\documentclass[pra,superscriptaddress,showpacs,twocolumn,10pt]{revtex4}
\usepackage{amssymb}
\usepackage{graphicx}

\def\>{\rangle}

\begin{document}
\newtheorem{corollary}{Corollary}
\newtheorem{definition}{Definition}
\newtheorem{example}{Example}
\newtheorem{lemma}{Lemma}
\newtheorem{proposition}{Proposition}
\newtheorem{statement}{Statement}
\newtheorem{theorem}{Theorem}
\newtheorem{fact}{Fact}
\title{Trade-off between multiple-copy transformation and entanglement catalysis}
\author{Runyao Duan}
\email{dry02@mails.tsinghua.edu.cn}
\author{Yuan Feng}
\email{feng-y@tsinghua.edu.cn}
\author{Xin Li}
\email{x-li02@mails.tsinghua.edu.cn}
\author{Mingsheng Ying}
\email{yingmsh@tsinghua.edu.cn} \affiliation{State Key Laboratory
of Intelligent Technology and Systems, Department of Computer
Science and Technology, Tsinghua University, Beijing, China,
100084}
\date{\today}
\begin{abstract}
We demonstrate that multiple copies of a bipartite entangled pure
state may serve as a catalyst for certain entanglement
transformations while a single copy cannot. Such a state is termed
a ``multiple-copy catalyst" for the transformations. A trade-off
between the number of copies of source state and that of the
catalyst is also observed. These results can be generalized to
probabilistic entanglement transformations directly.
\end{abstract}
 \maketitle

\section{Introduction}

In recent years, more and more applications of quantum information
processing, such as quantum cryptography \cite{BB84}, quantum
superdense coding \cite{BS92}, and quantum teleportation
\cite{BBC+93}, have led us to view quantum entanglement as a new
kind of physical resource \cite{M00}.  One of the central problems
about quantum entanglement is to find the conditions under which
an entangled state could be converted into another one  by using
local quantum operations and classical communication (LOCC) only.
Bennett and his collaborators \cite{BBPS96,Bet+96,BDAW96} made
significant progress in attacking this challenging problem for the
asymptotic setting, while for the deterministic transformations,
the first important step was made by Nielsen in Ref. \cite{NI99},
where he found a necessary and sufficient condition for a
bipartite entangled pure state to be transformed to another pure
one deterministically, under the constraint of LOCC. More
precisely, suppose that Alice and Bob share an entangled state
$|\psi\rangle$, and they want to transform it into another state
$|\varphi\rangle$ by using only local quantum operations on their
own subsystems and classical communication between them. Nielsen
proved that the two parties can finish this task successfully,
i.e., transforming $|\psi\rangle$ to $|\varphi\rangle$ with
certainty under LOCC, if and only if ${\psi}\prec{\varphi}$, where
$\psi$ and $\varphi$ denote the Schmidt coefficient vectors of
$|\psi\rangle$ and $|\varphi\rangle$, respectively. Here the
symbol `$\prec$' stands for `majorization relation', which is a
vast topic in linear algebra. For details about majorization,
please see Refs. \cite{MO79} and \cite{AU82}.

Nielsen's result implies that there can be two entangled pure
states, say $|\psi\rangle$ and $|\varphi\rangle$, such that they
are incomparable in the sense that neither the transformation of
$|\psi\rangle$ to $|\varphi\rangle$ nor the transformation of
$|\varphi\rangle$ to $|\psi\rangle$ can be realized with
certainty. For transformations between incomparable states, Vidal
\cite{Vidal99} generalized Nielsen's result to a probabilistic
version and established an explicit expression of the maximal
conversion probability for the transformation of $|\psi\rangle$ to
$|\varphi\rangle$ under LOCC.

Shortly after Nielsen's work, a quite surprising phenomenon of
entanglement, namely, entanglement catalysis, or ELOCC, was
discovered by Jonathan and Plenio~\cite{JP99}. They demonstrated
by examples  that sometimes one may use an entangled state
$|\phi\rangle$, known as a catalyst, to make an impossible
transformation of $|\psi\rangle$ to $|\varphi\rangle$ possible.
That is, in the presence of $|\phi\rangle$,  the transformation of
$|\psi\rangle\otimes|\phi\rangle$ to
$|\varphi\rangle\otimes|\phi\rangle$ can be realized with
certainty, which means that the catalyst $|\phi\rangle$ is not
modified in the process. A concrete example is as follows. Take
$|\psi\rangle=\sqrt{0.4}|00\rangle+\sqrt{0.4}|11\rangle+\sqrt{0.1}|22\rangle+\sqrt{0.1}|33\rangle$
and
$|\varphi\rangle=\sqrt{0.5}|00\rangle+\sqrt{0.25}|11\rangle+\sqrt{0.25}|22\rangle$.
We know that $|\psi\rangle$ cannot be transformed to
$|\varphi\rangle$ with certainty under LOCC but if another
entangled state
$|\phi\rangle=\sqrt{0.6}|44\rangle+\sqrt{0.4}|55\rangle$ is
introduced, then the transformation of
$|\psi\rangle\otimes|\phi\rangle$ to
$|\varphi\rangle\otimes|\phi\rangle$ can be realized with
certainty because $\psi\otimes \phi\prec \varphi\otimes \phi$. The
role of the state $|\phi\rangle$ in this transformation is similar
to a catalyst in a chemical process since it can help the
entanglement transformation process without being consumed. In the
same paper, Jonathan and Plenio also showed that the use of
catalyst can improve the maximal conversion probability when the
transformation cannot realize with certainty even with the help of
a catalyst. The mathematical structure of entanglement catalysis
was thoroughly studied in Ref. \cite{DK01}.

Bandyopadhyay $et\ al$ found another interesting phenomenon
\cite{SRS02}:  sometimes multiple copies of the source state may
be transformed into the same number of copies of the target state
although the transformation cannot happen for a single copy. Such
a phenomenon is called `nonasymptotic bipartite pure-state
entanglement transformation' in Ref. \cite{SRS02}. More
intuitively, this phenomenon can also be called `multiple-copy
entanglement transformation', or MLOCC for short. Take the above
states $|\psi\rangle$ and $|\varphi\rangle$ as an example. It is
not difficult to check that the transformation of
$|\psi\rangle^{\otimes 3}$ to $|\varphi\rangle^{\otimes 3}$ occurs
with certainty by Nielsen's theorem. That is, when Alice and Bob
prepare three copies of $|\psi\rangle$ instead of just a single
one, they can transform these three copies all together into the
same number of copies of $|\varphi\rangle$ by LOCC. This simple
example means that the effect of catalyst can, at least in the
above situation, be implemented by preparing a sufficiently large
number of copies of the original state and transforming these
copies together. Some important aspects of MLOCC were investigated
in Ref. \cite{SRS02}.

In this paper we examine the catalysis power when multiple copies
of catalyst state are available. What was discovered by
Bandyopadhyay \textit{et al} is that sometimes the effect of
catalysis can be implemented by increasing the number of copies of
the source state, whereas we present some examples to show another
interesting phenomenon:  a large enough number of copies of an
entangled pure state may act as a catalyst although a single copy
cannot. Such an entangled pure state can be called a
`\textit{multiple-copy catalyst}'. More formally, if
$|\phi\rangle$ is not a catalyst for the transformation of
$|\psi\rangle$ to $|\varphi\rangle$, but there is an integer $m>1$
such that $|\phi\rangle^{\otimes m}$ is a catalyst for the same
transformation, then $|\phi\rangle$ is called a multiple-copy
catalyst for the transformation of $|\psi\rangle$ to
$|\varphi\rangle$. A necessary condition for a given entangled
pure state to be a multiple-copy catalyst for a specific
transformation is obtained.

It is worth noting that both  ways of enabling entanglement
transformations in Ref. \cite{SRS02} and in the present paper are
increasing the number of the copies of states. The essential
difference is that in Ref. \cite{SRS02} the number of copies of
the source state is increased while in this paper we consider
increase the copies of catalysts. A lot of heuristic examples lead
us to find a trade-off between the number of copies of the
original entangled state and that of the catalyst. As is expected,
the more original-state copies are provided, the less catalyst
copies are needed and vice versa.

A similar phenomenon  also exists in the case of probabilistic
entanglement transformations. We show by examples that sometimes
the combination of MLOCC and ELOCC can increase the maximal
conversion probability efficiently.  We also present a necessary
condition for when the combination of multiple-copy
transformations and entanglement-assisted transformations has
advantages over pure LOCC transformations.

The rest of the paper is organized as follows. In Sec. II, we
study the combination of MLOCC and ELOCC in deterministic
transformations. These results are generalized  to probabilistic
ones in Sec. III. The paper is concluded in Sec. IV with some open
problems that may be of interest for the further study.

\section{Combining  MLOCC with ELOCC: deterministic case}
In this section, we give some examples to show that sometimes the
role of catalysts can be implemented by  multiple copies of
catalysts.

For the sake of convenience, we present here Nielsen's theorem
\cite{NI99} as a lemma since it will be used frequently to analyze
the possibility of entanglement transformations latter:

\begin{lemma}\label{lemma1}\upshape
Let $|\psi\rangle=\sum_{i=1}^{n}\sqrt{\alpha_i}|i\rangle|i\rangle$
and
$|\varphi\rangle=\sum_{i=1}^{n}\sqrt{\beta_i}|i\rangle|i\rangle$
be pure bipartite states with the Schmidt  coefficient vectors
$\psi=(\alpha_1, \ldots,\alpha_n)$ and
$\varphi=(\beta_1,\ldots,\beta_n)$, where
$\alpha_1\geq\cdots\geq\alpha_n \geq0$ and
$\beta_1\geq\cdots\geq\beta_n\geq 0$. Then there exists a
transformation that converts $|\psi\rangle$ into $|\varphi\rangle$
with certainty under LOCC if and only if  $\psi\prec \varphi$,
i.e.,
\begin{equation}\label{nielsen}
\sum_{i=1}^{l}\alpha_i\leq\sum_{i=1}^{l}\beta_i, {\ }1\leq l\leq
n,
\end{equation}
with equality when $l=n$.
\end{lemma}

Nielsen's theorem establishes an connection between the
transformation of $|\psi\rangle$ to $|\varphi\rangle$ and the
mathematical relation $\psi\prec \varphi$. Intuitively, we often
write $|\psi\rangle \prec |\varphi\rangle$ instead of $\psi\prec
\varphi$. From that one can immediately deduce that the
transformation of $|\psi\rangle$ to $|\varphi\rangle$ can be
achieved with certainty under LOCC.

As a useful application of Nielsen's theorem, we present a
technical lemma as follows:
\begin{lemma}\label{lemma2}\upshape
Let $|\psi\rangle$ and $|\varphi\rangle$ be two bipartite
entangled pure states. If $|\psi \rangle^{\otimes p} \prec
|\varphi\rangle^{\otimes p}$ for each $p=k,k+1,\ldots,2k-1$, then
$|\psi\rangle ^{\otimes p} \prec|\varphi\rangle^{\otimes p}$ for
all $p\geq k$.
\end{lemma}
In other words,  to check whether $|\psi\rangle^{\otimes p}\prec
|\varphi\rangle^{\otimes p}$ holds for every $p\geq k$, one only
needs to check $k$ values of $p$, i.e., $p=k,\ldots, 2k-1.$

{\bf Proof.} By Nielsen's theorem and the assumptions, to prove
that $|\psi\rangle^{\otimes p}\prec |\varphi\rangle^{\otimes p}$
for every $p\geq k$, we only need to show that the transformation
of $|\psi\rangle^{\otimes p}$ to $|\varphi\rangle^{\otimes p}$ can
be realized with certainty for any $p\geq 2k$.  For this purpose,
we uniquely decompose the positive integer $p$ such that $p\geq
2k$ as
\begin{equation}\label{pdecom}
p=(r-1)k+ (k+s),  {\rm \ \ }r\geq 2 {\rm \ and\ } 0\leq s\leq k-1.
\end{equation}
Now an explicit protocol implementing the transformation  of
$|\psi\rangle ^{\otimes p}$ to $|\varphi\rangle^{\otimes p}$ with
certainty  under LOCC consists of the following two steps:

1). Perform $(r-1)$ times of the transformation of $|\psi\rangle
^{\otimes k}$ to $|\varphi\rangle^{\otimes k}$;

2). Perform one time of the transformation of $|\psi\rangle
^{\otimes k+s}$ to $|\varphi\rangle^{\otimes k+s}.$

By  Nielsen's theorem and the assumptions again,  we know that
both the transformations in 1) and 2)  can be realized with
certainty by LOCC. That completes the proof of Lemma \ref{lemma2}. \hfill $\blacksquare$\\

It is worth noting that the conditions in Lemma \ref{lemma2} are
also necessary in general. In fact, as pointed out by Leung and
Smolin in Ref. \cite{LS01}, the majorization relation is not
monotonic in general in the sense that $|\psi\rangle^{\otimes
k}\prec |\varphi\rangle^{\otimes k}$ does not always imply
$|\psi\rangle^{\otimes k+1}\prec |\varphi\rangle^{\otimes k+1}$.
Thus, to guarantee that $|\psi\rangle^{\otimes p}\prec
|\varphi\rangle^{\otimes p}$ holds for every $p\geq k$, one needs
to check all $k$ conditions.

Now we begin to examine the catalysis power when multiple copies
of catalyst state are available. In particular, the following
example indicates the existence of multiple-copy catalyst.

\begin{example}\label{example1}\upshape
Suppose that the original entangled state owned by Alice and Bob
is
\begin{equation}\label{source1}
|\psi\rangle=\sqrt{0.4}|00\rangle+\sqrt{0.4}|11\rangle+\sqrt{0.1}|22\rangle+\sqrt{0.1}|33\rangle,
\end{equation}
and the final state they want to transform $|\psi\rangle$ into is
\begin{equation}\label{target1}
|\varphi\rangle=\sqrt{0.5}|00\rangle+\sqrt{0.25}|11\rangle+\sqrt{0.22}|22\rangle+\sqrt{0.03}|33\rangle.
\end{equation}

This example is very close to the original one  used by Jonathan
and Plenio \cite{JP99} to demonstrate the effect of catalysis. One
may think that Alice and Bob could realize the transformation of
$|\psi\rangle$ to $|\varphi\rangle$ with a $2\times 2$ catalyst,
as in the original example in Ref. \cite{JP99}. Unfortunately, it
is not the case since the small deviation violates the condition
of the existence of a $2\times 2$ catalyst \cite{SD03}. However,
we can find a $3\times 3$ state
\begin{equation}\label{3catalyst}
|\phi_1\rangle=\sqrt{\frac{50}{103}}|44\rangle+\sqrt{\frac{30}{103}}|55\rangle+\sqrt{\frac{23}{103}}|66\rangle
\end{equation}
such that $|\psi\rangle\otimes|\phi_1\rangle\prec
|\varphi\rangle\otimes|\phi_1\rangle$.

Moreover, by a routine calculation, we may observe that
\begin{equation}
|\psi\rangle^{\otimes k}\nprec |\varphi\rangle^{\otimes k},\ \
1\leq k\leq 4,
\end{equation}
but
\begin{equation}\label{13}
|\psi\rangle^{\otimes k}\prec |\varphi\rangle^{\otimes k},\ \
5\leq k\leq 9
\end{equation}
holds. Thus Eq. (\ref{13}) is true for any $k\geq 5$ by Lemma
\ref{lemma2}. Again, this shows that the effect of a catalyst can
be implemented by increasing the number of copies of the source
state in a transformation. We now further put
\begin{equation}\label{2catalyst1}
|\phi_2\rangle=\sqrt{0.6}|44\rangle+\sqrt{0.4}|55\rangle,
\end{equation}
which is certainly not a catalyst for the transformation mentioned
above. An interesting thing here is that $|\phi_2\rangle^{\otimes
5}$ does serve as a catalyst for the transformation of
$|\psi\rangle$ to $|\varphi\rangle$ because an easy calculation
shows that $|\psi\rangle\otimes|\phi_2\rangle^{\otimes 5
}\prec|\varphi\rangle\otimes|\phi_2\rangle^{\otimes 5}$. Of
course, $|\phi_2\rangle^{\otimes 5}$ is not the optimal one in the
sense that its dimension is not the minimum among all catalysts.
This phenomenon indicates that increasing the number of an
entangled
pure state may strictly broaden the power of its catalysis.\hfill $\blacksquare$\\
\end{example}

In the next example, we combine MLOCC with ELOCC, and show that a
tradeoff exists between the number of copies of source state and
that of catalyst.

\begin{example}\label{example2}\upshape
Suppose that Alice and Bob share some copies of source state
$|\psi\rangle$ as in Eq. (\ref{source1}), and they want to
transform it to the same number of copies of
\begin{equation}\label{target2}
|\varphi\rangle=\sqrt{0.5}|00\rangle+\sqrt{0.25}|11\rangle+\sqrt{0.2}|22\rangle+\sqrt{0.05}|33\rangle
\end{equation}
by LOCC. Suppose that the only states they can borrow from a
catalyst banker are some copies of $|\phi_2\rangle$ in Eq.
(\ref{2catalyst1}). Can Alice and Bob realize their task? Notice
that
\begin{equation}
|\psi\rangle^{\otimes 5} \not \prec |\varphi\rangle^{\otimes
 5}\ {\rm but }\ |\psi\rangle^{\otimes k} \prec |\varphi\rangle^{\otimes
 k}, {\rm\ }6\leq k \leq 11.
\end{equation}
Applying Lemma \ref{lemma2} yields that if the number of available
copies of $|\psi\rangle$ is larger than or equal to 6, then Alice
and Bob always can realize their task by themselves without
borrowing any catalyst. But if they only own 5  copies of
$|\psi\rangle$, they cannot realize the transformation even if
joint operations on the 5 copies are performed.  It is easy to
check that borrowing one copy of $|\phi_2\rangle$ is enough for
Alice and Bob's task because $|\psi\rangle^{\otimes 5}\otimes
|\phi_2\rangle\prec |\varphi\rangle^{\otimes 5}\otimes
|\phi_2\rangle.$ Similarly, when they only own 4 copies of
$|\psi\rangle$, it is sufficient to finish the task successfully
by borrowing 2 copies of $|\phi_2\rangle$. For the case that 3
copies of $|\psi\rangle$ are owned by Alice and Bob, it is easy to
see that 3 copies of $|\phi_2\rangle$ are not enough for their
purpose and the minimal number of $|\phi_2\rangle$ is 4. Finally,
when Alice and Bob own only one copy of $|\psi\rangle$, using $6$
to $10$ copies of $|\phi_2\rangle$ cannot achieve the task. We
conclude that they must borrow at leat 11 copies of
$|\phi_2\rangle$ from the catalyst banker since the relation
$|\psi\rangle\otimes|\phi_2\rangle^{\otimes k}\prec
|\varphi\rangle\otimes |\phi_2\rangle^{\otimes k}$ holds only for
$k\geq 11$. Here we have used Nielsen's theorem and the fact that
if $|\phi\rangle^{\otimes k}$ is a catalyst for the transformation
of $|\psi\rangle$ to $|\varphi\rangle$ then $|\phi\rangle^{\otimes
p}$ is also a catalyst for the same transformation for any $p\geq
k$. Alice and Bob must borrow a large number of catalysts to
complete the transformation in this extreme case.  This example
illustrates a tradeoff between the number of copies of original
state and that of catalyst.\hfill $\blacksquare$\\
\end{example}

The above two examples show that it will be very useful to know
when a given entangled pure state can serve as a multiple-copy
catalyst for a specific entanglement  transformation.
Unfortunately, such a characterization is not known at present.
Nevertheless, we can give a necessary condition for the existence
of multiple-copy catalyst.

Before presenting this necessary condition, we introduce some
useful notations. We define $x^{\downarrow}$ as the vector which
is obtained by rearranging the components of $x$ into the
nonincreasing order. A useful fact about this notation is that
$x^{\downarrow}=y^{\downarrow}$ if and only if the components of
$x$ are exactly the same as those of $y$. In other words, they are
equivalent up to a permutation. For any bipartite entangled pure
states $|\psi\rangle$ and $|\varphi\rangle$ with the ordered
Schmidt coefficient vectors $\psi^{\downarrow}=(\alpha_1,\ldots,
\alpha_n)$ and $\varphi^{\downarrow}=(\beta_1,\ldots,\beta_n)$, we
define a set of indices as
\begin{equation}\label{defnonequality}
L_{\psi, \varphi}=\{l:1\leq l< n\ {\rm and}\ \sum_{j=1}^l \alpha_j
> \sum_{j=1}^l \beta_j\}.
\end{equation}
Intuitively, for any $l\in L_{\psi, \varphi}$, the sum of the $l$
largest components of $\psi$ is strictly larger than that of
$\varphi$. So $|\psi\rangle$ and $|\varphi\rangle$ are
incomparable if and only if $L_{\psi,\varphi}\neq \emptyset$ and
$L_{\varphi,\psi}\neq \emptyset$.

The following lemma is interesting in its own right. It gives us a
necessary condition for a bipartite  entangled pure state
$|\phi\rangle$ with Schmidt coefficients $\gamma_1\geq
\gamma_2\geq \cdots \geq\gamma_k> 0$ to be a  catalyst for a given
transformation.

\begin{lemma}\label{lemma3}\upshape
Let $|\psi\rangle$ and $|\varphi\rangle$ be two incomparable
states. If $|\phi\rangle$ is a catalyst for the transformation of
$|\psi\rangle$ to $|\varphi\rangle$, then for any $l\in L_{\psi,
\varphi}$, it holds that
${\gamma_1}/{\gamma_k}>{\beta_l}/{\beta_{l+1}},$ and
\begin{equation}\label{equ1}
\frac{\gamma_1}{\gamma_i}>\frac{\beta_{l}}{\beta_{l+1}} {\rm \ \ \
or\ \ \ }\frac{\gamma_{i}}{\gamma_{i+1}}<\frac{\beta_{1}}{\beta_l}
\end{equation}
and
\begin{equation}\label{equ2}
\frac{\gamma_{i+1}}{\gamma_{k}}>\frac{\beta_{l}}{\beta_{l+1}} {\rm
\ \ \ or\ \ \
}\frac{\gamma_{i}}{\gamma_{i+1}}<\frac{\beta_{l+1}}{\beta_{n}}.
\end{equation}
for $i=1,\ldots, k-1$.
\end{lemma}

{\bf Proof.} By contradiction, suppose that  one of the following
holds:

Case (a): there exist $l_0\in L_{\psi,\varphi}$ and $1\leq i_0\leq
k-1$ such that either Eq. (\ref{equ1}) or Eq. (\ref{equ2}) does
not hold;

Case (b): there exists $l_0\in L_{\psi,\varphi}$ such that
${\gamma_1}/{\gamma_k}\leq{\beta_{l_0}}/{\beta_{{l_0}+1}}$.

We only need to  prove that both Cases (a) and (b) contradict the
assumption $|\psi\rangle\otimes |\phi\rangle\prec
|\varphi\rangle\otimes \phi\rangle$.

Firstly, we deal with Case (a). Let us decompose $\psi$ into two
shorter vectors $\psi'$ and $\psi''$, that is, $\psi=(\psi',
\psi'')$, such that $\psi'=(\alpha_1,\ldots,\alpha_{l_0})$ and
$\psi''=(\alpha_{{l_0}+1},\ldots,\alpha_n)$. $\varphi$ is
similarly decomposed as $\varphi=(\varphi',\varphi'')$. We also
decompose $\phi=(\phi', \phi'')$, where
$\phi'=(\gamma_1,\ldots,\gamma_{i_0})$ and
$\phi''=(\gamma_{i_0+1},\ldots,\gamma_k)$.

Since $\varphi\otimes \phi=(\varphi',\varphi'')\otimes
(\phi',\phi'')$, one can easily check that the components of
$\varphi\otimes \phi$ are exactly the same as those of
$(\varphi'\otimes \phi', \varphi'\otimes \phi'',\varphi''\otimes
\phi',\varphi''\otimes \phi'')$ by a simple algebraic calculation.
By our notations introduced above, we always have
\begin{equation}\label{decompose}
(\varphi\otimes \phi)^{\downarrow}=(\varphi'\otimes \phi',
\varphi'\otimes \phi'',\varphi''\otimes \phi',\varphi''\otimes
\phi'')^{\downarrow}.
\end{equation}
Notice that the minimal component of $\varphi'\otimes \phi'$ is
$\beta_{l_0}\gamma_{i_0}$, while the maximal components of
$\varphi'\otimes \phi''$, $\varphi''\otimes \phi'$, and
$\varphi''\otimes \psi''$ are $\beta_1\gamma_{i_0+1}$,
$\beta_{l_0+1}\gamma_1$, and $\beta_{l_0+1}\gamma_{i_0+1}$,
respectively.

To finish the proof of Case (a), it suffices to consider the
following two subcases:

Subcase (a.1): Eq. (\ref{equ1}) is not satisfied, that is,
\begin{equation}\label{nequ1}
{\gamma_1}/{\gamma_{i_0}}\leq {\beta_{l_0}}/{\beta_{l_0+1}} {\rm \
\ and\
 \ } {\gamma_{i_0}}/{\gamma_{i_0+1}}\geq{\beta_{1}}/{\beta_{l_0}},
\end{equation}

then
\begin{equation}\label{max}
\beta_{l_0}\gamma_{i_0} \geq \max\{\beta_1\gamma_{i_0+1},
\beta_{l_0+1}\gamma_1, \beta_{l_0+1}\gamma_{i_0+1}\},
\end{equation}
which implies that the minimal component  of $\varphi'\otimes
\phi'$ is not less than  the maximal components of
$\varphi'\otimes \phi''$, $\varphi''\otimes \phi'$ and
$\varphi''\otimes \phi''$. By Eqs. (\ref{decompose}) and
(\ref{max}), the largest $i_0l_0$ components of $\varphi\otimes
\phi$ are just the components of $\varphi'\otimes \phi'$. So
\begin{equation}\label{contradiction}
\begin{array}{rl}
\displaystyle\sum_{j=1}^{i_0l_0}(\varphi\otimes \phi)^\downarrow_j
&=\displaystyle\sum_{j=1}^{i_0l_0} (\varphi'\otimes
\phi')^{\downarrow}_j\\
&= (\displaystyle\sum_{j=1}^{l_0}
\beta_j) (\displaystyle\sum_{j=1}^{i_0} \gamma_j) \\
 &< \displaystyle(\sum_{j=1}^{l_0} \alpha_j) (\sum_{j=1}^{i_0}
\gamma_j)\\
&=\displaystyle\sum_{j=1}^{i_0l_0} (\psi'\otimes
\phi')^{\downarrow}_j\\
&\leq \displaystyle\sum_{j=1}^{i_0l_0}(\psi\otimes
\phi)^\downarrow_j,
\end{array}
\end{equation}
where the strict inequality follows  from $l_0\in
L_{\psi,\varphi}$, while the last inequality is by the definition
of $\sum_{j=1}^{i_0l_0}(\psi\otimes \phi)^\downarrow_j$. It
follows that $|\psi\rangle\otimes |\phi\rangle\nprec
|\varphi\rangle\otimes |\phi\rangle$, a contradiction.

Subcase (a.2):  Eq. (\ref{equ2}) is not satisfied. Then  by
similar arguments we can verify that the least
$(k-{i_0})(n-{l_0})$ components of $\varphi\otimes \phi$ are just
the components of $\varphi''\otimes \phi''$, and thus
$|\psi\rangle\otimes |\phi\rangle \nprec |\varphi\rangle\otimes
|\phi\rangle$ by considering the sum of the least
$(k-{i_0})(n-{l_0})$ components of $\varphi\otimes \phi$. This is
also a contradiction.

Now we deal with the Case (b). In this case, $\phi'=\phi$ and
$\phi''$ disappears. With almost the  same arguments as in Subcase
(a.1), we have that  $|\psi\rangle\otimes |\phi\rangle\nprec
|\varphi\rangle\otimes |\phi\rangle$, again a contradiction. That
completes the proof of Lemma \ref{lemma3}. \hfill $\blacksquare$

In the above lemma, if we take $i=1$ then from Eq. (\ref{equ1}) we
have $\gamma_{1}/\gamma_{2}<\beta_{1}/\beta_l$. Similarly, taking
$i=k-1$ leads us to $\gamma_{k-1}/\gamma_{k}<\beta_{l+1}/\beta_n$
from Eq. (\ref{equ2}). Consequently, we have the following
corollary:
\begin{corollary}\label{corollary2}\upshape
Let $|\psi\rangle$ and $|\varphi\rangle$ be two incomparable
states. If $|\phi\rangle$ is a catalyst for the transformation of
$|\psi\rangle$ to $|\varphi\rangle$, then for any $l\in L_{\psi,
\varphi}$,
\begin{equation}\label{conditionnoncompelte}
\frac{\gamma_{1}}{\gamma_{2}}<\frac{\beta_{1}}{\beta_l}\ \ {\rm\
and\ }\ \
\frac{\gamma_{k-1}}{\gamma_{k}}<\frac{\beta_{l+1}}{\beta_{n}}.
\end{equation}
\end{corollary}

The following theorem indicates that the condition in Eq.
(\ref{conditionnoncompelte}) is also necessary for $|\phi\rangle$
to be a multiple-copy catalyst for the transformation of
$|\psi\rangle$ to $|\varphi\rangle$.

\begin{theorem}\label{theorem1}\upshape
Let $|\psi\rangle$ and $|\varphi\rangle$ be two incomparable
states. If $|\phi\rangle$ is a multiple-copy catalyst for the
transformation of $|\psi\rangle$ to $|\varphi\rangle$, then for
any $l\in L_{\psi,\varphi}$, Eq. (\ref{conditionnoncompelte})
holds.
\end{theorem}

{\bf Proof.} If $|\phi\rangle$ is a multiple-copy catalyst for the
transformation  of $|\psi\rangle$ to $|\varphi\rangle$, then there
exists a positive integer $m$ such that $|\phi\rangle^{\otimes m}$
is a catalyst the same transformation. By Corollary
\ref{corollary2}, it follows that

\begin{equation}\label{m1}
\frac{{(\phi^{\otimes m})^\downarrow_1}}{{(\phi^{\otimes
m})^\downarrow_2}}<\frac{\beta_1}{\beta_l}
\end{equation}
and
\begin{equation}\label{m2}
\frac{(\phi^{\otimes m})^\downarrow_{k^m-1}}{(\phi^{\otimes
m})^\downarrow_{k^m}}<\frac{\beta_{l+1}}{\beta_n}
\end{equation}
for any $l\in L_{\psi,\varphi}$.

It is easy to check that
\begin{equation}\label{m3}
\frac{{(\phi^{\otimes m})^\downarrow_1}}{{(\phi^{\otimes
m})^\downarrow_2}} = \frac{\gamma_1^m}{\gamma_2\gamma_1^{m-1}}=
\frac{\gamma_1}{\gamma_2}
\end{equation}
and
\begin{equation}\label{m4}
\frac{(\phi^{\otimes m})^\downarrow_{k^m-1}}{(\phi^{\otimes
m})^\downarrow_{k^m}} =
\frac{\gamma_k^{m-1}\gamma_{k-1}}{\gamma_k^m}=\frac
{\gamma_{k-1}}{\gamma_{k}}.
\end{equation}
Combining  Eqs. (\ref{m1})--(\ref{m4}), we  have the validity of
Eq. (\ref{conditionnoncompelte}). This completes the proof of
Theorem \ref{theorem1}.\hfill $\blacksquare$\\

With the help of Theorem 1, we are able to find a state
$|\phi\rangle$ such that it is a multiple-copy catalyst for the
transformation of $|\psi\rangle^{\otimes k}$ to
$|\varphi\rangle^{\otimes k}$ with some $k>1$, but not for the
transformation of $|\psi\rangle$ to $|\varphi\rangle$.
Intuitively, multiple-copy transformation can be catalyzed more
easily than single-copy transformation.

\begin{example}\label{example3}\upshape
Take the source state as
\begin{equation}\label{source3}
|\psi'\rangle=\frac{1}{\sqrt{1.01}}(|\psi\rangle+\sqrt{0.01}|44\rangle),
\end{equation}
while the target as
\begin{equation}\label{target3}
|\varphi'\rangle=\frac{1}{\sqrt{1.01}}(|\varphi\rangle+\sqrt{0.01}|44\rangle),
\end{equation}
where $|\psi\rangle$ and $|\varphi\rangle$ are defined as Eq.
(\ref{source1}) and Eq. (\ref{target2}), respectively. We choose
\begin{equation}\label{extended}
|\phi_3\rangle=\sqrt{0.7}|55\rangle+\sqrt{0.3}|66\rangle.
\end{equation}
A simple calculation shows that $|\phi_3\rangle$ is a catalyst for
5-copy transformation ( i.e., the transformation of
$|\psi'\rangle^{\otimes 5} $ to $|\varphi'\rangle^{\otimes 5} $),
and $|\phi_3\rangle^{\otimes 2}$ is a catalyst both for 4-copy
transformation and for 3-copy transformation. It is obvious that
$L_{\psi',\varphi'}=\{2\}$,
$\varphi'=\frac{1}{1.01}(0.5,0.25,0.2,0.05,0.01)$ and
$\phi_3=(0.7,0.3)$. So
$$\frac{\gamma_1}{\gamma_2}=\frac{0.7}{0.3}>\frac{0.5}{0.25}=\frac{\beta_1}{\beta_2},$$
which yields that the condition in Eq.
(\ref{conditionnoncompelte}) is violated. Thus by Theorem
\ref{theorem1},  it follows that $|\phi_3\rangle$ is not a
multiple-copy catalyst for the transformation of $|\psi'\rangle$
to $|\varphi'\rangle$. In other words, for arbitrarily large $q$,
the transformation of $|\psi'\rangle\otimes
|\phi_3\rangle^{\otimes q}$ to $|\varphi'\rangle\otimes
|\phi_3\rangle^{\otimes q}$
cannot be achieved with certainty.\hfill $\blacksquare$\\
\end{example}

\section{Combining  MLOCC with  ELOCC: probabilistic case}

We concerned ourselves with deterministic transformations in the
last section. In this section, let us turn now to examine
entanglement transformations with probability strictly less than
$1$.

Recall Vidal's theorem from Ref. \cite{Vidal99} that the maximal
conversion probability of  transforming $|\psi\rangle$ to
$|\varphi\rangle$ under LOCC is given by

\begin{equation}\label{vidal}
P_{max}(|\psi\rangle\rightarrow|\varphi\rangle)={\rm min}_{1\leq
l\leq n} \frac{E_l(|\psi\rangle)}{E_l(|\varphi\rangle)},
\end{equation}
where $E_l(|\psi\rangle)=\sum_{i=l}^n\alpha_i$ and $\alpha_1\geq
\alpha_2\geq \cdots\geq\alpha_n$ are the Schmidt coefficients of
$|\psi\rangle$.

Let $\lambda\in(0,1]$. We call $|\phi\rangle$ a
\textit{$\lambda$-catalyst} for the transformation of
$|\psi\rangle$ to $|\varphi\rangle$ if

\begin{equation}P_{max}(|\psi\rangle\otimes |\phi\rangle \rightarrow
|\varphi\rangle\otimes |\phi\rangle) \geq \lambda.
\end{equation}
Furthermore, if $|\phi\rangle^{\otimes k}$ serves as a
$\lambda$-catayst for some integer $k>1$, then we say that
$|\phi\rangle$ is a \textit{multiple-copy $\lambda$-catalyst} for
this transformation.

We say that a transformation $|\psi\rangle$ of $|\varphi\rangle$
can attain probability $\lambda$ by MLOCC if there exists a
positive integer $k$ such that
\begin{equation}P_{max}(|\psi\rangle^{\otimes k}\rightarrow
|\varphi\rangle^{\otimes k}) \geq \lambda^k.
\end{equation}
Notice that if the maximal conversion probability from
$|\psi\rangle$ to $|\varphi\rangle$ by LOCC is $\lambda$, then the
right-hand side of the above equation is just the maximal
conversion probability of transforming $|\psi\rangle^{\otimes k}$
into $|\varphi\rangle^{\otimes k}$ separately, that is, in a way
where no collective operations on the $k$ copies are performed.
Thus the intuition behind the above definition is that with the
help of MLOCC, the average probability of a single-copy
transformation is not less than $\lambda$.

With the above preliminaries, the results obtained in Section II
can be directly extended into the probabilistic case. The
following example, first considered by Jonathan and Plenio in
\cite{JP99}, demonstrates the existence of multiple-copy
$\lambda$-catalysts. It also shows that the presence of
multiple-copy $\lambda$-catalyst and multiple copies of source
state can increase the maximal conversion probability efficiently.

\begin{example}\label{example4}\upshape
Let
$|\psi\rangle=\sqrt{0.6}|00\rangle+\sqrt{0.2}|11\rangle+\sqrt{0.2}|22\rangle$
and
$|\varphi\rangle=\sqrt{0.5}|00\rangle+\sqrt{0.4}|11\rangle+\sqrt{0.1}|22\rangle$.
By Vidal's theorem, we have that
$P_{max}(|\psi\rangle\rightarrow|\varphi\rangle)=0.80$. However,
with the aid of an entangled state
$|\phi\rangle=\sqrt{0.65}|33\rangle+\sqrt{0.35}|44\rangle$, the
maximal conversion probability becomes
$P_{max}(|\psi\rangle\otimes
|\phi\rangle\rightarrow|\varphi\rangle\otimes|\phi\rangle)=0.904$,
which means that $|\phi\rangle$ is a $0.904$-catalyst for the
transformation of $|\psi\rangle$ to $|\varphi\rangle$. Can Alice
and Bob increase their conversion probability to $0.985$?  A
careful analysis shows that the transformation of $|\psi\rangle$
to $|\varphi\rangle$ does not have any $2\times 2$
$0.985$-catalyst \cite{DFY04}. Fortunately, $|\phi\rangle$ is a
multiple-copy $0.985$-catalyst since
\begin{equation}
P_{max}(|\psi\rangle\otimes |\phi\rangle^{\otimes 19}\rightarrow|
\varphi\rangle\otimes|\phi\rangle^{\otimes 19})\geq 0.985.
\end{equation}

Suppose now that Alice and Bob share two copies of $|\psi\rangle$.
According to our definition, the transformation of $|\psi\rangle$
to $|\varphi\rangle$ can attain a probability
$(0.8533)^{1/2}=0.9237$ under MLOCC since
\begin{equation}\label{2prob}
P_{max}(|\psi\rangle^{\otimes 2}
\rightarrow|\varphi\rangle^{\otimes 2})=0.8533.
\end{equation}

If we combine catalyst-assisted transformation and multiple-copy
one together, the maximal conversion probability can increase
efficiently. For example,
\begin{equation}
P_{max}(|\psi\rangle^{\otimes 2}\otimes |\phi\rangle^{\otimes3}
\rightarrow|\varphi\rangle^{\otimes 2}\otimes|\phi\rangle^{\otimes
3 })=0.9535
\end{equation}
This implies that the transformation of $|\psi\rangle$ to
$|\varphi\rangle$ can attain the probability $0.9535^{1/2}=0.9765$
under the combination of MLOCC and ELOCC. In contrast to that, a
pure MLOCC needs at least $7$ copies of
$|\psi\rangle$ to attain the  probability $0.985$.\hfill $\blacksquare$\\
\end{example}

Next, let us turn to another interesting question: is it always
useful to combine catalyst-assisted transformation with
multiple-copy transformation? The above two examples give some
hints to a positive answer to the question. However, the next
theorem indicates that such an improvement does not always happen.
This theorem is a generalization of Lemma 4 in \cite{JP99} which
says that the presence of catalysts cannot always increase
conversion probability. We should point out that a similar result
has also been obtained in \cite{SRS02}.

For any bipartite entangled pure states $|\psi\rangle$ and
$|\varphi\rangle$, we define
\begin{equation}\label{eq1}
P_{max}^{E}(|\psi\rangle\rightarrow | \varphi\rangle)= {\rm
sup}_{|\phi\rangle}P_{max}(|\psi\rangle\otimes
|\phi\rangle\rightarrow|\varphi\rangle\otimes|\phi\rangle).
\end{equation}
Intuitively, $P_{max}^{E}(|\psi\rangle\rightarrow |
\varphi\rangle)$ denotes the optimal conversion probability of
transforming $|\psi\rangle$ to $|\varphi\rangle$ by using some
catalyst.

\begin{theorem}\label{theorem2}\upshape
 Let $|\psi\rangle$ and
$|\varphi\rangle$ be two $n\times n$ states with the least Schmidt
coefficients  $\alpha_n$ and $\beta_n$, respectively. Then we have
that
\begin{equation}\label{probcatalyst}
(P_{max}(|\psi\rangle\rightarrow |\varphi\rangle))^p\leq
P_{max}^E(|\psi\rangle^{\otimes
p}\rightarrow|\varphi\rangle^{\otimes p})\leq
(\frac{\alpha_n}{\beta_n})^p
\end{equation}
for any positive integer $p$.
\end{theorem}

{\bf Proof.} The first inequality in Eq. (\ref{probcatalyst}) is
obtained by performing the transformation of
$|\psi\rangle^{\otimes p}$ to $|\varphi\rangle^{\otimes p}$ under
LOCC separately. The second inequality in Eq. (\ref{probcatalyst})
can be proven as follows. Suppose that $|\phi\rangle$ is any
entangled pure state with the least Schmidt coefficient $\gamma_k>
0$. By Vidal's theorem, we obtain that
\begin{equation}\label{eq2}
\begin{array}{l} P_{max}(|\psi\rangle^{\otimes p }\otimes
|\phi\rangle\rightarrow|\varphi\rangle^{\otimes
p}\otimes|\phi\rangle) \\
\\
=\displaystyle{\rm min}_{1\leq l\leq n^{p}k
}\frac{E_l(|\psi\rangle^{\otimes p }\otimes
|\phi\rangle)}{E_l(|\varphi\rangle^{\otimes p }\otimes
|\phi\rangle)}\\
\\
\leq \displaystyle\frac{E_{n^{p}k}(|\psi\rangle^{\otimes p
}\otimes |\phi\rangle)}{E_{n^{p}k}(|\varphi\rangle^{\otimes p
}\otimes
|\phi\rangle)}=\frac{\alpha_{n}^{p}\gamma_{k}}{\beta_{n}^{p}\gamma_{k}}
=(\frac{\alpha_{n}}{\beta_{n}})^p,
\end{array}
\end{equation}
where we have used the fact that $E_{n^{p}k}(|\psi\rangle^{\otimes
p }\otimes |\phi\rangle)=\alpha_{n}^{p}\gamma_{k}$. The second
inequality of Eq. (\ref{probcatalyst}) follows from Eqs.
(\ref{eq1}) and (\ref{eq2}). This
completes the proof of the theorem. \hfill$\blacksquare$\\

\begin{corollary}\label{corollary1}\upshape
With the same assumption as in Theorem 2, if
$P_{max}(|\psi\rangle\rightarrow
|\varphi\rangle)=\frac{\alpha_{n}}{\beta_{n}}$, then
$P_{max}^E(|\psi\rangle^{\otimes p}\rightarrow |
\varphi\rangle^{\otimes p})= (\frac{\alpha_n}{\beta_n})^p$.
\end{corollary}
In other words, even the combination of multiple-copy
transformation and catalyst-assisted transformation cannot
increase the conversion probability. In fact, collective
operations in this case have no advantages over individual
operations.

An interesting application of Corollary \ref{corollary1} is to
deal with the case when $|\varphi\rangle$ is a maximally entangled
state, that is, $|\varphi\rangle=\frac{1}{\sqrt{n}}\sum_{i=1}^n
|i\rangle|i\rangle$. The maximal conversion probability
$P_{max}(|\psi\rangle\rightarrow |\varphi\rangle)=n\alpha_n$
cannot be increased by any combination of multiple-copy
transformations and entanglement-assisted  ones. Example
\ref{example4} gives another application of the corollary. In
fact, for any $3\times 3$ dimensional $|\psi\rangle$ and
$|\varphi\rangle$, if $\alpha_3< \beta_3$, then it follows from
Vidal's theorem that $P_{max}(|\psi\rangle\rightarrow
|\varphi\rangle)=\frac{\alpha_3}{\beta_3}$. Hence by the above
corollary, $P^E_{max}(|\psi\rangle^{\otimes p} \rightarrow
|\varphi\rangle^{\otimes p})=(\frac{\alpha_3}{\beta_3})^p$, which
is exponentially decreasing when $p$ increases, as pointed out in
Ref. \cite{SRS02}.

\section{conclusion}
To summarize, we have demonstrated that in some cases multiple
copies of an entangled state can serve as a catalyst although a
single copy cannot. Such a state is called a `multiple-copy
catalyst'. We have analyzed the power of combining MLOCC with
ELOCC. Moreover, a tradeoff between the number of copies of source
state and that of catalyst is observed. We also show that the
combination of MLOCC and ELOCC can increase the maximal conversion
probability efficiently. Note that there are no analytical ways
for finding a catalyst for a given transformation except for some
special cases \cite{SD03}\cite{DFLY04}. The notion of
multiple-copy catalyst sometimes may lead us to a possible way of
seeking an intended catalyst.

There are many open problems that may be of relevance. The most
interesting one is, of course, what is the precise relation
between MLOCC and ELOCC? Furthermore, is the combination of MLOCC
and ELOCC always more powerful than separate MLOCC or ELOCC
\cite{DFLY04}? The another interesting one is to give a sufficient
condition for a given entangled state to be a  multiple-copy
catalyst for a certain transformation.

\smallskip\

\textbf{Acknowledgement:}  We thank Somshubhro Bandyopadhyay for
informing us valuable references and Lisha Huang for helpful
discussions about majorization. The acknowledgement is also given
to the colleagues in the Quantum Computation and Quantum
Information Research Group, especially to Zhengfeng Ji for many
inspiring discussions about MLOCC. This work was partly supported
by the Natural Science Foundation of China (Grant Nos. 60273003,
60433050, 60321002, and 60305005). Runyao Duan acknowledges the
financial support of Tsinghua University (Grant No. 052420003).
\smallskip\


\begin{thebibliography}{9999}
\bibitem{DFY04}   In fact,  we can calculate the maximal conversion probability
of transforming  $|\psi\rangle$ to $|\varphi\rangle$ with the aid
of a $2\times 2$ catalyst is $\frac{5+\sqrt{985}}{40}\thickapprox
90.96\%$.
\end{thebibliography}
\end{document}